\documentclass[a4paper,10pt]{article}
\usepackage[dvips]{graphicx}
\usepackage{amsfonts,amsmath,amssymb}

\begin{document}
\title{\bf Neutrino mixing matrix and masses from a generalized
Friedberg-Lee model }
\author{N. Razzaghi \footnote{Electronic address: neda.phys@gmail.com} \quad and
\quad S. S. Gousheh \footnote{Electronic address: ss-gousheh@sbu.ac.ir} \\
\small Department of Physics, Shahid Beheshti University, G. C.
Evin, Tehran 19839 Iran}

\maketitle

\begin{abstract}
 The overall characteristics of the solar and atmospheric neutrino
 oscillation are approximately consistent with a tribimaximal
form of the mixing matrix $U$ of the lepton sector. Exact
tribimaximal mixing leads to $\theta_{13}=0$. However, recent
results from Daya Bay and RENO experiments have established a
nonzero value for $\theta_{13}$. Keeping the leading behavior of
$U$ as tribimaximal, we use a generalized Fridberg-Lee neutrino
mass model along with a complementary ansatz to incorporate a
nonzero $\theta_{13}$ along with CP violation. We generalize this
model in two stages: In the first stage we assume $\mu-\tau$
symmetry and add imaginary components which leads to nonzero
phases. In the second stage we add a perturbation with real
components which breaks the $\mu-\tau$ symmetry and this leads to
a nonzero value for $\theta_{13}$. The combination of these two
generalizations leads to CP violation. Using only two of the
experimental data, we can fix all of the parameters of our model
and predict not only values for the other experimental data, which
agree well with the available data, but also the masses of
neutrinos and the CP violating phases and parameters. These
predictions include the following: $\langle m_{\nu_e}
\rangle\approx(0.033-0.037)~eV$, $\langle m_{\nu_\mu}
\rangle\approx(0.043-0.048)~ eV$, $\langle m_{\nu_\tau}
\rangle\approx(0.046-0.051)~ eV$, and $59.21^{\circ}\lesssim
\delta\lesssim 59.34^{\circ}$
\end{abstract}

\section{Introduction}\label{sec1}
The remarkable experimental achievements resulting in experimental
data for solar, atmospheric, accelerator and reactor neutrinos
\cite{exp} give us information about neutrino masses and mixing
which can be summarized as follows in Table \ref{h1}
\cite{information1, Valle}:
\begin{center}
\begin{table}[h1]
\centering
\begin{tabular}{|c|c|c|}
\hline
Parameter &  The experimental data     & The best fit ($\pm1\sigma$)\\
\hline
$\Delta m_{21}^{2}[10^{-5}eV^{2}]$ & $7.12-8.20$ & $7.43-7.81$  \\
\hline
$|\Delta m_{31}^{2}|[10^{-3}eV^{2}]$& $2.13-2.74$ & $2.46-2.61$ \\
~&$2.21-2.64$&$2.37-2.50$ \\
 \hline
$\sin^{2}\theta_{12}$  & $0.27-0.37$ & $0.303-0.336$  \\
\hline
$\sin^{2}\theta_{13}$& $0.017-0.033$ & $0.0218-0.0275$\\
~&~&$0.0223-0.0276$ \\
\hline
$\sin^{2}\theta_{23}$& $0.36-0.68$ & $0.400-0.461 ~\text{and}~ 0.573-0.635$  \\
~&$0.37-0.67$&$0.569-0.626$\\
 \hline
$\delta$& $0-2\pi$ & $0-2\pi$ \\
\hline
\end{tabular}
\caption{The experimental data for the neutrinos' mixing
parameters. In the last column, when multiple sets of allowed
ranges are stated, the upper row corresponds to normal hierarchy
and the lower raw to inverted hierarchy.}\label{h1}
\label{tab:PPer}
\end{table}
\end{center}

The lepton mixing matrix in the standard parametrization is given
by \cite{mixing,para},

\begin{equation}\label{emixing}
U_{PMNS}=\left(\begin{array}{ccc}c_{12}c_{13} & s_{12}c_{13} & s_{13}e^{-i\delta}\\
-s_{12}c_{23}-c_{12}s_{23}s_{13}e^{i\delta} &
c_{12}c_{23}-s_{12}s_{23}s_{13}e^{i\delta} &
s_{23}c_{13}\\s_{12}s_{23}-c_{12}c_{23}s_{13}e^{i\delta}
& -c_{12}s_{23}-s_{12}c_{23}s_{13}e^{i\delta} & c_{23}c_{13}\end{array}\right)\left(\begin{array}{ccc}1 & 0 & 0 \\
0 & e^{i\rho} & 0\\0 & 0 & e^{i\sigma}\end{array}\right),
\end{equation}
where $c_{ij}\equiv\cos\theta_{ij}\text{ and
}s_{ij}\equiv\sin\theta_{ij}$ (for $i,j=(1,2), (1,3) \text{ and } (2,3) $).
The phase $ \delta $ is called the Dirac phase, analogous to the
CKM phase, and the phases $\rho$ and $ \sigma$ are called the
Majorana phases and are relevant for Majorana neutrinos. However,
we should mention that recently the advantages of the original
symmetrical form of the parameterizations of the lepton mixing
matrix has been discussed \cite{para}.

A relatively successful phenomenological Ansats for the neutrino
mixing matrix was proposed by Harrison, Perkins and Scott in 2002
\cite{Harrison}, and is known as the tribimaximal mixing
matrix (TBM). It has the following form,

\begin{equation}\label{etbm}
U_{TBM} =\left(\begin{array}{ccc}-\sqrt{\frac{2}{3}} & \frac{1}{\sqrt{3}} & 0\\
\frac{1}{\sqrt{6}} & \frac{1}{\sqrt{3}} &
-\frac{1}{\sqrt{2}}\\\frac{1}{\sqrt{6}} & \frac{1}{\sqrt{3}} &
\frac{1}{\sqrt{2}}\end{array}\right).
\end{equation}
In this matrix the third mixing angle, {\em i.e.} $\theta_{13}$, is exactly
zero. However the results of the Daya Bay and RENO collaborations
have shown that $\theta_{13}=0$ is now rejected at a significance
level higher than $8\sigma$. A combined analysis of the data coming from
T2K, MINOS, Double Chooz, and Daya Bay experiments shows that the
best-fit value of $\theta_{13}$ is given by
$\sin^{2}\theta_{13}=0.026(0.027)_{-0.004}^{+0.003}$ for normal
(or inverted) mass hierarchy. From these significant data, it has to be concluded
that the simple-picture of tribimaximal mixing matrix fails.
However the smallness of $\theta_{13}$ as compared to the other
two mixing angles encourages us to examine whether a small
perturbation on the basic tribimaximal structure could lead to a
nonzero value for $\theta_{13}$ and a more realistic neutrino
mixing matrix.

A successful phenomenological neutrino mass model with flavor
symmetry that is suitable for both the Dirac and the Majorana neutrinos was proposed
by Friedberg and Lee (FL). This model is interesting
because when $\mu-\tau$ symmetry is assumed, the resulting
neutrino mixing matrix reduces to the $U_{TBM}$. Usually the
ansatz for the  neutrino mass matrices contain more parameters
than can be measured in realistic experiments. Thus, the presence
of certain conditions or simplifications for the neutrino mass
matrix is useful. What might first come to mind is the presence of
zeros in the mass matrix \cite{zeromass}. However, conditions on
basis independent quantities, namely the trace and the determinant, are
more suitable to consider. The simplest conditions are to set these quantities to zero. The
condition $det (M_{\nu})=0$ \cite{det0} leads to zero mass for one of the neutrinos.
A zero determinant can be motivated on various
grounds \cite{det00}. The second simplest basis independent
condition is a vanishing trace, {\em i.e.}, $Tr (M_{\nu})=0$,
which is called the complementary ansatz. Its consequences have
first been investigated in \cite{trc0} applying a three neutrino
framework that simultaneously explains the anomalies of solar and
atmospheric neutrino oscillation experiments as well as the LSND
experiment. In \cite{trc01} a CP conserving traceless $M_{\nu}$
has been investigated for the simpler case of explaining only the
atmospheric and solar neutrino deficits. Motivations for traceless
mass matrices can be provided by models in which $M_{\nu}$ is
constructed through a commutator of two matrices, as it happens in
models of radiative mass generation \cite{trc02}. More
interestingly, an approximately traceless
$M_{\nu}$ can be the consequence of exact $b-\tau$ unification at
high energy scales within type II see-saw models \cite{trc03},
which in this framework is also the reason for maximal atmospheric
neutrino mixing \cite{trc04,trc05}. The type II see-saw mechanism
was the original motivation of the traceless $M_{\nu}$ condition
as investigated in \cite{trc0}. Also in \cite{trc06} the condition
$Tr (M_{\nu})=0$ has been used to determine the values of neutrino
masses, and the CP phases when CP violation is considered. In the
present paper we will use the FL model as a starting point, and
assume the complementary ansatz.

In the FL model the mass eigenstates of three charged leptons are
identified with their flavor eigenstates. Therefore neutrino
mixing matrix can be simply described by $3\times3$ unitary matrix
$U$ which transforms the neutrino mass eigenstates to the flavor
eigenstates, $ (\nu_{e},\nu_{\mu},\nu_{\tau}).$ The Dirac
neutrino mass operator in the FL model can be written as
\begin{eqnarray}\label{emfl}\vspace{.5cm}
  {\cal{M}}_{FL}&=&a\left(\bar{\nu}_{\tau}-\bar{\nu}_{\mu}\right)\left(\nu_{\tau}-\nu_{\mu}\right)
+
b\left(\bar{\nu}_{\mu}-\bar{\nu}_{e}\right)\left(\nu_{\mu}-\nu_{e}\right)
+ c\left(\bar{\nu}_{e}-\bar{\nu}_{\tau}\right)\left(\nu_{e}-\nu_{\tau}\right)\nonumber\\
&+&
m_{0}\left(\bar{\nu}_{e}\nu_{e}+\bar{\nu}_{\mu}\nu_{\mu}+\bar{\nu}_{\tau}\nu_{\tau}\right).
\end{eqnarray}
All the parameters in this model ($a,b,c$ and $m_{0}$) are assumed
to be real. In the original FL model, also known as the pure FL model,
$m_0=0$ and in this case $M_{FL}$ has the following
symmetry $\nu_{e}\rightarrow\nu_{e}+z $,
$\nu_{\mu}\rightarrow\nu_{\mu}+z $, and $
\nu_{\tau}\rightarrow\nu_{\tau}+z $, where $z$ is an element of
the Grassman algebra. For constant $z$, this symmetry is called
the FL symmetry \cite{FL} in which case the kinetic term is also
invariant. However the other terms of the electroweak Lagrangian
do not have such a symmetry. The $ m_{0} $ term  breaks this
symmetry explicitly. However we may add that the FL symmetry leads
to a magic matrix and this property is not spoiled by the $m_{0}$
term. The magic property has many manifestations which we shall
discus in detail. Also it has been reasoned that the FL symmetry
is the residual symmetry of the neutrino mass matrix after the
$SO(3)\times U(1)$ flavor symmetry breaking \cite{FL2}. The mass
matrix can be displayed as,
\begin{equation}\label{efl}
M_{FL} =\left(\begin{array}{ccc}b+c+m_{0} & -b & -c\\
-b & a+b+m_{0} & -a\\-c & -a & a+c+m_{0}\end{array}\right),
\end{equation}
where $a \propto\left(Y_{\mu\tau}+Y_{\tau\mu }\right)$, $b
\propto\left(Y_{e\tau}+Y_{\tau e }\right)$ and $  \ c
\propto\left(Y_{\tau e}+Y_{e\tau }\right) $ and $Y_{\alpha\beta}$
denote the Yukawa coupling constants. The proportionality constant
is the expectation value of the Higgs field. It is apparent that $
M_{FL} $ possesses exact $\mu-\tau$ symmetry only when $b=c$.
Setting $b=c$ and using the hermiticity of $M_{FL}$, a straight
forward diagonalization procedure yields $U^{T}M_{FL}U=\text{ Diag
} \{m^{(0)}_{1},m^{(0)}_{2},m^{(0)}_{3}\} $ where,
\begin{eqnarray}\label{emm1}\vspace{.2cm}
m^{(0)}_{1}&=&3b+m_{0}\nonumber\\m^{(0)}_{2}&=&m_{0},~~~~~~~~~~~~~~~~~\text{and}~~~~~~U=U_{TBM},\nonumber
\\m^{(0)}_{3}&=&2a+b+m_{0}.
\end{eqnarray}

Note that in the pure FL
model one of the neutrino masses is exactly zero.
For a general exact TBM neutrino mixing, regardless of the model,
the mixing angles are given by $\sin^2\theta_{12}=\frac{1}{3}$
($\theta_{12} \approx35.3^{\circ}$),
$\sin^2\theta_{23}=\frac{1}{2}$ ($\theta_{23}=45^{\circ}$),
$\theta_{13}=0$ and the CP-violating phases can be chosen to be
zero. In order to have CP-violation in the standard parametrization
given in Eq.\,(\ref{emixing}), the necessary
condition is $\delta\neq0$ and $\theta_{13}\neq0$. In this model
these conditions necessarily mandate that $\mu-\tau$ symmetry
should be broken. Another interesting question is whether
$\theta_{23}=45^{\circ}$ holds after the $\mu-\tau$
symmetry breaking.

There are four independent CP-even quadratic invariants, which can
conveniently be chosen as $U^{\ast}_{11}U_{11},
U^{\ast}_{13}U_{13}, U^{\ast}_{21}U_{21}$ and $
U^{\ast}_{23}U_{23} $ and three independent CP-odd quartic
invariants \cite{quadratic},
\begin{eqnarray}\label{eJ}\vspace{.2cm}
J&=&{\cal{I}}m(U_{11}U^{\ast}_{12}U^{\ast}_{21}U_{22})\nonumber\\
I_{1}&=&{\cal{I}}m[(U^{\ast}_{11}U_{12})^{2}]\nonumber\\I_{2}&=&{\cal{I}}m[(U^{\ast}_{11}U_{13})^{2}].
\end{eqnarray}
The Jarlskog rephasing invariant parameter $J$ \cite{J}, is relevant for CP
violation in lepton number conserving processes like neutrino
oscillations. $I_{1}$ and $I_{2}$ are relevant for CP violation in
lepton number violating processes like neutrinoless double beta
decay. Oscillation experiments cannot distinguish between the
Dirac and Majorana neutrinos. The detection of neutrinoless double
beta decay would provide direct evidence of lepton number
non-conservation and the Majorana nature of neutrinos. Many
theoretical and phenomenological works have discussed  massive
neutrino models that break $\mu-\tau$ symmetry as a prelude to CP
violation\cite{theoretical}.

In this paper we start with the simple FL model with exact
$\mu-\tau$ symmetry, which leads to $U_{TBM}$ mixing matrix. We
then generalize this model by adding complex parameters to the
elements of $M_{FL}$ which can ultimately be linked to complex
Yukawa coupling constants. This generalization can be broken down
to two clearly distinguishable pieces. In this paper we first
study the results of each piece separately and then investigate
the results when both pieces are applied simultaneously. Both
pieces are separately exact solvable, while the combination is
not. First we add a real term proportional to the matrix
representation of the permutation group element $S^{132}$, merely
to break the $\mu-\tau$ symmetry. As we shall show, it suffices to
treat this piece as a perturbation. We find that this generates a
nonzero $\theta_{13}$ whose smallness justifies the use of
perturbation theory. Second we add arbitrary imaginary
coefficients while preserving $\mu-\tau$ symmetry. This results in
the generation of nonzero phases, including $\delta$, which turn
out to be large. For combining the two steps, we first do the
complexification step exactly and finally add the $\mu-\tau$
symmetry breaking perturbation. We then find that nonzero
values are generated for both $\theta_{13}$ and $\delta$, and this leads to CP
violation.

This paper is organized as follows. In section 2,
we introduce our model and show the results of the two aforementioned generalization,
separately. We then combine the two generalization and show the results. We use some
self consistency arguments along with the complementary ansatz to reduce the number
of our free parameters. In section 3, we map two of the experimental data onto the
allowed region of our parameter space. We then check the consistency of all other
experimental data with the two overlap regions, which selects only one region.
This region is extremely small and almost pinpoints all of our parameters.
The results indicate that the complexification had to be treated nonperturbatively,
since the phases turn out to be large. However the coefficient for $\mu-\tau$
symmetry breaking justifies the use of perturbation theory for that part, since $\theta_{13}$
turns out to be small. We then not only check the consistency of all of our
results with the available experimental data, but also present our predictions
for the actual masses and CP violation parameters. In section 4, we state our conclusions.

\section{The Model}\label{sec2}
In this section, we study the effects of generalizations of the FL model that break
the $\mu-\tau$ and CP symmetries. In the first
stage we perturb the mass matrix so as to break the $\mu-\tau$
symmetry. We choose the mass matrix to be of the form
$M_\nu=M_\nu^{0}+M_\nu^{'}=M_\nu^{0}+\lambda S^i$, where
$M_\nu^{0}=M_{FL}$ with $b=c$, $\lambda$ is a perturbation
parameter with dimension of mass, and $S^i$ is an element of the
permutation matrices which breaks the $\mu-\tau$ symmetry. For
example for $S^i=S^{(132)}$, we have,

\begin{eqnarray}\label{em1}
M_{\nu} &=&M_{\nu}^{0}+M_{\nu}^{'}=M_\nu^{0}+\lambda
S^{(132)}\nonumber
\\&=&\left(\begin{array}{ccc}2b+m_{0} & -b & -b\\
-b & a+b+m_{0} & -a\\-b & -a & a+b+m_{0}\end{array}\right)+\lambda\left(\begin{array}{ccc}0 & 1 & 0\\
0 & 0 & 1\\1& 0 & 0\end{array}\right),
\end{eqnarray}
Notice that $M_{\nu}$ and $M_{FL}$ are both magic and symmetric
matrices since they both commute with the magic $S$ matrix defined
by
\begin{equation}\label{es}
S =\left(\begin{array}{ccc}F & T & T\\
T & F & T\\T & T & F\end{array}\right).
\end{equation}
Therefore one of the eigenstates must be $(\frac{1}{\sqrt{3}} ,
\frac{1}{\sqrt{3}} , \frac{1}{\sqrt{3}})$. The elements of the basis
in which $M_{\nu}^{0}$ is diagonal are simply the columns of
$U_{TBM}$, as shown in Eq.\,(\ref{etbm}), and are as follows,
\begin{equation} \vspace{.2cm}\label{evm01}
\nu^{(0)}_{1}=\left(\begin{array}{ccc}-\sqrt{\frac{2}{3}}\\
\frac{1}{\sqrt{6}}\\\frac{1}{\sqrt{6}}
\end{array}\right), ~~~~\nu^{(0)}_{2}=\left(\begin{array}{ccc}\frac{1}{\sqrt{3}}\\
\frac{1}{\sqrt{3}}\\\frac{1}{\sqrt{3}}\end{array}\right), ~~~~~\nu^{(0)}_{3}=\left(\begin{array}{ccc}0 \\
-\frac{1}{\sqrt{2}}\\\frac{1}{\sqrt{2}}\end{array}\right)
.
\end{equation}
Here $M_{\nu}^{0}$ and $M_{\nu}^{'}$ are real matrices. Needless
to say, this may generate a non-zero $\theta_{13}$ but will not
lead to CP violation, since this computation necessarily yields
$\delta=0$. As stated in the Introduction, although this problem
is exactly solvable it not only suffices to treat this part as a
perturbation since $\theta_{13}$ turns out to be small, but also
it is advantageous to do so for comparison purposes with the part
when all generalizations are combined. The perturbation expansion
of the mass eigenstates of $M_\nu$ to first order is
 \begin{equation}\label{ep1}
\nu_{i}=\nu_{i}^{(0)}+\sum_{i\neq
j}\frac{\langle\nu_{j}^{(0)}|M_{\nu}^{'}|\nu_{i}^{(0)}\rangle}{m_{i}^{0}-
m_{j}^{0}}\nu_{j}^{(0)},
\end{equation}
The $\nu_{i}$s
are simply the columns of the mixing matrix $U$ and are as follows,

\begin{equation}\label{evm1}
\nu_{1}=\left(\begin{array}{ccc}-\sqrt{\frac{2}{3}}\\
\frac{1}{\sqrt{6}}-\frac{\sqrt{\frac{3}{32}}\lambda}{(a-b)}\\\frac{1}{\sqrt{6}}+\frac{\sqrt{\frac{3}{32}}\lambda}{(a-b)}
\end{array}\right),~~~~\nu_{2}=\left(\begin{array}{ccc}\frac{1}{\sqrt{3}}\\
\frac{1}{\sqrt{3}}\\\frac{1}{\sqrt{3}}\end{array}\right),~~~~\nu_{3}=\left(\begin{array}{ccc}-\frac{\lambda}{2\sqrt{2}(a-b)} \\
-\frac{1}{\sqrt{2}}+\frac{\lambda}{4\sqrt{2}(a-b)}\\\frac{1}{\sqrt{2}}+\frac{\lambda}{4\sqrt{2}(a-b)}\end{array}\right).
\end{equation}

Comparing $\nu_{3}$ given in Eq.\,(\ref{evm1}) with the third
column of $U_{PMNS}$ as given by  Eq.\,(\ref{emixing}), we immediately obtain all of the
mixing angles ($ \theta_{13} , \theta_{12}, \theta_{23} $) and the
CP violating phase $\delta$ in terms of the parameters of our
model, {\em i.e.} $a, b$ and $\lambda$, as follows,
\begin{eqnarray}\label{esin}
\sin^{2}\theta_{13}&=&\frac{2\lambda^{2}}{16(a-b)^{2}+3\lambda^{2}},
\nonumber\\\sin^{2}\theta_{12}&=&\frac{1}{3\cos^{2}\theta_{13}}=\frac{16(a-b)^{2}
+3\lambda^{2}}{48(a-b)^{2}+3\lambda^{2}},
\nonumber\\\sin^{2}\theta_{23}&=&\frac{1}{2}-\frac{4\lambda(a-b)}{16(a-b)^{2}+\lambda^{2}},
\nonumber\\\tan\delta&=&0.
\end{eqnarray}
Notice that we managed to obtain a nonzero $\theta_{13}$.

The first order corrections to the neutrino masses are obtained from
$\langle\nu_{i}^{(0)}|M_{\nu}^{'}|\nu_{j}^{(0)}\rangle=
m_{i}^{(1)}\delta_{ij}$. Therefore the masses up to first order
correction are as follows,
\begin{equation}\label{eme1}\vspace{.2cm}
m^{(\lambda)}_{1}=3b+m_{0}-\frac{\lambda}{2},~~~~~~~~ m^{(\lambda)}_{2}=
m_{0}+\lambda,~~~~~~~~m^{(\lambda)}_{3}=2a+b+m_{0}-\frac{\lambda}{2}.
\end{equation}
From the complementary ansatz we obtain,
$a=\frac{-(4b+3m_{0})}{2}$. In the limit $\lambda\rightarrow0$,
the mixing matrix $U$ must reduce to $U_{TBM}$. Therefore we find
$(a-b)<0$, which has also been obtained in \cite{me}. Using these
conditions, along with the usual convention of FL model in which
$m_0>0$, one can easily show that $-\frac{1}{2} m_0\lesssim
b\lesssim 0$, $-\frac{3}{2} m_0\lesssim a\lesssim -\frac{1}{2}
m_0$, $m_3<0$, $m_3<m_1$ and $\Delta m^2_{31}=m_3^2-m_1^2>0$.
Therefore, we have normal hierarchy at this stage. As we shall see
this property remains true in all stages of our model.

In the second stage we let all of the coefficients ${a, b, c}$ of
the original $M_{FL}$ be complex with the restriction $b=c$ which
preserves the exact $\mu-\tau$ symmetry of the original
$M_\nu^{0}$. This mass matrix can be displayed as
 \begin{equation}\label{em02}
M^{c0}_{\nu} =\left(\begin{array}{ccc}2(b+iB)+m_{0} & -(b+iB) & -(b+iB)\\
-(b+iB) & (a+iA)+(b+iB)+m_{0} & -(a+iA)\\-(b+iB) & -(a+iA) &
(a+iA)+(b+iB)+m_{0}\end{array}\right),
\end{equation}
Where the superscript ``c'' stands for ``complex''. The
$M^{c0}_{\nu}$ is a non-Hermitian matrix, so in general  we need
two distinct unitary matrices $U$ and $V$ to diagonalize it, {\em
i.e.} $M^{c0}_{diag}=U^{\dag}M^{C0}_{\nu}V$. These matrices can be
easily obtained by diagonalizing $
M^{c0}_{\nu}{M^{c0}_{\nu}}^{\dag} $ and $
{M^{c0}_{\nu}}^{\dag}M^{c0}_{\nu} $, separately. The matrices $U$
and $V$ are the conventional transformation matrices for the
left-handed and right-handed neutrinos, respectively. An
interesting point is that for ${M^{c0}_{\nu}}$ one can easily show
that $U=V=U_{TBM}$, due to the $\mu-\tau$ symmetry. Therefore the
basis in which ${M^{c0}_{\nu}}$ is diagonal is the same as that of
${M^{0}_{\nu}}$, {\em i.e.} $\{\nu_{i}^{(0)},~i=1, 2, 3\}$ given by
Eq.\,(\ref{evm01}). The eigenvalues of $M^{c0}_{\nu}$ are as
follows,

\begin{equation}\label{eme2}\vspace{.2cm}
m^{c0}_{1}=(3b+m_{0})+i(3B),~~~~~ m^{c0}_{2}=
m_{0},~~~~~m^{c0}_{3}=(2a+b+m_{0})+i(2A+B).
\end{equation}
From the complementary ansatz, {\em i.e.} $Tr(M^{c0}_{\nu})=0$, we
obtain,
\begin{equation}\label{eAa}\vspace{.2cm}
a=\frac{-(4b+3m_{0})}{2},~~\text{and}~~A=-2B.
\end{equation}
It is worth mentioning that in this case because of the imaginary
terms in the mass matrix $M^{c0}_{\nu}$, we have two phases that
appear in the mass eigenvalues shown in Eq.\,(\ref{eme2}), {\em
i.e.} $\exp[i \arctan({{3B}\over{3b+m_0}})]$ and $\exp[i
\arctan({{2A+B}\over{2a+b+m_0}})]$. For the Dirac neutrinos these
phases can be removed and for the Majorana neutrinos these phases
remain as Majorana phases and contribute to CP violation.

Notice that for the complex FL model with $\mu-\tau$ symmetry $\theta_{13}=0$. Moreover, the
``zero sum'' condition, {\em i.e.} $m^{c0}_{1}+m^{c0}_{2}+m^{c0}_{3}=0$, is
equivalent to the traceless condition, since
$Tr(M^{c0}_{\nu})=Tr(U^{\dag}VM^{c0}_{diag})=Tr(M^{c0}_{diag})=0$.
However, notice that in general $Tr(M)\neq Tr(M_{diag})$
\cite{trc01}.

The experimental data have now definitely confirmed that $\Delta
m^2_{21}=m^2_2-m^2_1>0$. Therefore, using Eq.\,(\ref{eme2}), we
can define the following new variable,

\begin{equation}\label{eB}
B^{'} :=|y|\sqrt{-b^{'}\left(b^{'}+\frac{2}{3}\right)},
\end{equation}
 where $-1<y<1$ and we have defined dimensionless parameters: $b^{'}\equiv\frac{b}{m_{0}}$,
$B^{'}\equiv\frac{B}{m_{0}}$ and $a^{'}\equiv\frac{a}{m_{0}}$. The
limit $|y|\rightarrow 1$ leads to
$|m^{c0}_{2}|\rightarrow|m^{c0}_{1}|$. The conditions stated for
$a$ and $b$ below Eq.\,(\ref{eme1}), {\em i.e.}
$-\frac{1}{2}\lesssim b^{'}\lesssim 0$ and $-\frac{3}{2}\lesssim
a^{'}\lesssim-\frac{1}{2}$, are consistent with the definition of
$B^{'}$.

In the final stage we combine two basic generalization to the $M_{FL}$, namely the addition
of imaginary components while preserving $\mu-\tau$ symmetry as a finite distortion,
and the addition of a permutation matrix with a real coefficient to break the $\mu-\tau$
symmetry as a perturbation.
Then we obtain $M^{c}_{\nu}$ as follows,
\begin{eqnarray}\label{em2}
M^{c}_{\nu} &=&M_{\nu}^{c0}+M_{\nu}^{'}\nonumber
\\&=&\left(\begin{array}{ccc}2(b+iB)+m_{0} & -(b+iB) & -(b+iB)\\
-(b+iB) & (a+iA)+(b+iB)+m_{0} & -(a+iA)\\-(b+iB) & -(a+iA) &
(a+iA)+(b+iB)+m_{0}\end{array}\right)\nonumber
\\&+&\lambda\left(\begin{array}{ccc}0 & 1 & 0\\
0 & 0 & 1\\1& 0 & 0\end{array}\right),
\end{eqnarray}
where $M_{\nu}^{c0}$ is a complex symmetric non-Hermitian matrix
given in Eq.\,(\ref{em02}) and $M_{\nu}^{'}=\lambda S^{(132)}$, as
given by Eq.\,(\ref{em1}), is real matrix. The columns of the
mixing matrix $U$ in Eq.\,(\ref{emixing}) are eigenstates of
$M^{c^{\dagger}}_{\nu}M^{c}_{\nu}=
M_{\nu}^{c0^{\dagger}}M_{\nu}^{c0}+M_{\nu}^{c0^{\dagger}}M_{\nu}^{'}+M_{\nu}^{'^{\dagger}}M_{\nu}^{c0}$,
where we have dropped a term which is $\mathcal{O}(\lambda^{2})$.
To proceed, we recall that $M_{\nu}^{c0^{\dagger}}M_{\nu}^{c0}$ is
Hermitian and its eigenstates are the same as the columns that
produce $U_{TBM}$, and its eigenvalues are $|m_{1}^{(c0)}|^{2}$,
$|m_{2}^{(c0)}|^{2}$, and $|m_{3}^{(c0)}|^{2}$.
The basis in which $M_\nu^c$ is diagonal is given by,
\begin{equation}\label{ep2}
\nu^{c}_{i}=\nu_{i}^{(0)}+\sum_{i\neq
j}\frac{\langle\nu_{j}^{(0)}|M_{\nu}^{c0^{\dagger}}M_{\nu}^{'}+M_{\nu}
^{'^{\dagger}}M_{\nu}^{c0}|\nu_{i}^{(0)}\rangle}{(m_{i}^{c0})^{2}-
(m_{j}^{c0})^{2}}\nu_{j}^{(0)}.
\end{equation}
A straight forward calculation yields,
\begin{eqnarray} \vspace{.2cm}\label{emb1}
\nu_{1}^{c}&=&\frac{2\sqrt{3}}{\sqrt{12+
\frac{\left(36B^{2}+(3-2m_{0})^{2}\right)\lambda^{2}}{(1+2b)^{2}(3-2m_{0})^{2}}}}\left(\begin{array}{ccc}-\sqrt{\frac{2}{3}}\\
\frac{1}{\sqrt{6}}-\frac{(3-2m_{0}-i6B)\lambda}{2\sqrt{6}(1+2b)(-3+2m_{0})}\\
\frac{1}{\sqrt{6}}+\frac{(3-2m_{0}-i6B)\lambda}{2\sqrt{6}(1+2b)(-3+2m_{0})}\end{array}\right),
\nonumber \nu_{2}^{c}=\left(\begin{array}{ccc}\frac{1}{\sqrt{3}}\\
\frac{1}{\sqrt{3}}\\\frac{1}{\sqrt{3}}\end{array}\right), \nonumber\\
\nu_{3}^{c}&=&\left(\begin{array}{ccc}\frac{-\sqrt{\frac{2}{3}}(3-2m_{0}+i6B)\lambda}{(1+2b)(-3+2m_{0})\sqrt{12
+\frac{\left(36B^{2}+(3-2m_{0})^{2}\right)\lambda^{2}}{(1+2b)^{2}(3-2m_{0})^{2}}}}\\\frac{-\frac{1}{\sqrt{2}}
+\frac{(3-2m_{0}+i6B)\lambda}{6\sqrt{2}(1+2b)(-3+2m_{0})}}{\frac{1}{2}\sqrt{4+\frac{\left(36B^{2}+(3-2m_{0})^{2}\right)
\lambda^{2}}{(1+2b)^{2}(3-2m_{0})^{2}}}}\\\frac{\frac{1}{\sqrt{2}}
+\frac{(3-2m_{0}+i6B)\lambda}{6\sqrt{2}(1+2b)(-3+2m_{0})}}{\frac{1}{2}\sqrt{4+\frac{\left(36B^{2}+(3-2m_{0})^{2}\right)
\lambda^{2}}{(1+2b)^{2}(3-2m_{0})^{2}}}}\end{array}\right) .
\end{eqnarray}
These are also simply the columns of mixing matrix, $U$, that contain
CP violating phases. Comparing Eq.\,(\ref{emb1}) with Eq.\,(\ref{emixing}), we
immediately obtain all of the mixing angles ($ \theta_{13} ,
\theta_{12}, \theta_{23} $) and the CP-violating phase in terms of
$ b^{'}$, $B^{'}$ and $\lambda^{'}\equiv\frac{\lambda}{m_0}$ as follows,
\begin{eqnarray}\label{esin}
\sin^{2}\theta_{13}&=&\frac{2(1+36B^{'2})\lambda^{'2}}{36(1+2b^{'})^{2}+3(1+36B^{'2})\lambda^{'2}},
\nonumber\\\sin^{2}\theta_{12}&=&1-\frac{24(1+2b^{'})^{2}}{36(1+2b^{'})^{2}+(1+36B^{'2})\lambda^{'2}},
\nonumber\\\sin^{2}\theta_{23}&=&\frac{1}{2}+\frac{6(1+2b^{'})\lambda^{'}}{36(1+2b^{'})^{2}+(1+36B^{'2})\lambda^{'2}},
\nonumber\\\tan\delta&=&6B^{'}.
\end{eqnarray}

The first order corrections to the neutrino masses are obtained
from
$m_{i}^{(1)}\delta_{ij}=\langle\nu_{i}^{(0)}|M_{\nu}^{c0^{\dagger}}M_{\nu}^{'}+M_{\nu}
^{'^{\dagger}}M_{\nu}^{c0}|\nu_{j}^{(0)}\rangle $. Therefore by
using Eq.\,(\ref{eme2}) and the first order of neutrino mass
correction we have,
\begin{eqnarray}\label{eme3}\vspace{.2cm}
|m^{c}_{1}|^2&=&|m^{c0}_{1}|^2-(3b+m_{0})\lambda,
\nonumber\\
|m^{c}_{2}|^2&=&|m^{c0}_{2}|^2+2m_{0}\lambda,\nonumber
\\|m^{c}_{3}|^2&=&|m^{c0}_{3}|^2-(2a+b+m_{0})\lambda.
\end{eqnarray}
As in Eq.\,(\ref{eme2}), $m^{c}_{1}$ and $m^{c}_{3}$ are complex.
Extracting the real and imaginary components of the masses and
introducing them as two phases in accordance with Eq.\,(\ref{emixing}), we obtain
$\rho=-\arctan\left(\frac{3B}{3b+m_{0}-\frac{\lambda}{2}}\right)$and
$\sigma=-\arctan\left(\frac{3B}{3b+m_{0}-\frac{\lambda}{2}}\right)
+\arctan\left(\frac{2A+B}{2a+b+m_{0}-\frac{\lambda}{2}}\right)$.
Since $M_{\nu}^{c0}$ is a symmetric matrix, it could also be used
as a Majorana mass matrix, and these phases can be considered as
the Majorana phases. In the Dirac case these phases are
transferred to the mass eigenstates. However in the Majorana case
the phase factors remain and contribute to the CP violation. The
requirement $det(U)=1$ leads to $\rho+\sigma=0$ \cite{phase}. This
in turn implies the following,
\begin{equation}\label{elanda}
\lambda^{'} =~2\left(b^{'}+\sqrt{(1+2b^{'})^{2}+3B^{'2}}\right).
\end{equation}

Using Eq.\,(\ref{eme3}), we can write the neutrino mass
squared differences as follows,
\begin{eqnarray}\label{eme33}\vspace{.2cm}
\Delta
m_{21}^{2}&=&|m^{c0}_{2}|^2-|m^{c0}_{1}|^2+3\left(b+m_0\right)\lambda,~~~~~~~\nonumber\\
\Delta
m_{31}^{2}&=&|m^{c0}_{3}|^2-|m^{c0}_{1}|^2-2\left(a-b\right)\lambda,\nonumber\\
\Delta
m_{32}^{2}&=&|m^{c0}_{3}|^2-|m^{c0}_{2}|^2-\left(2a+b+3m_0\right)\lambda.
\end{eqnarray}
A rephasing-invariant measure of CP violation in neutrino
oscillation is the universal parameter $J$ \cite{J} given in
Eq.\,(\ref{eJ}), and it has a form which is independent of the
choice of the Dirac or Majorana neutrinos. Using Eq.\,(\ref{emb1})
the expression for $J$ simplifies to,
\begin{equation}\label{eJJ}
J =~-\frac{4\left(1+2b^{'}\right)
B^{'}\lambda^{'}}{12+48b^{'}\left(1+b^{'}\right)
+\left(1+36B^{'2}\right)\lambda^{'2}}.
\end{equation}
Notice that we have reduced the number of free parameters of our
models to just two, {\em i.e.} $b^{'}$ and $y$, using all the self
consistency arguments and the complementary ansatz.

\section{Comparison with experimental data}\label{sec3}
\begin{center}
\begin{figure}[th] \includegraphics[width=12cm]{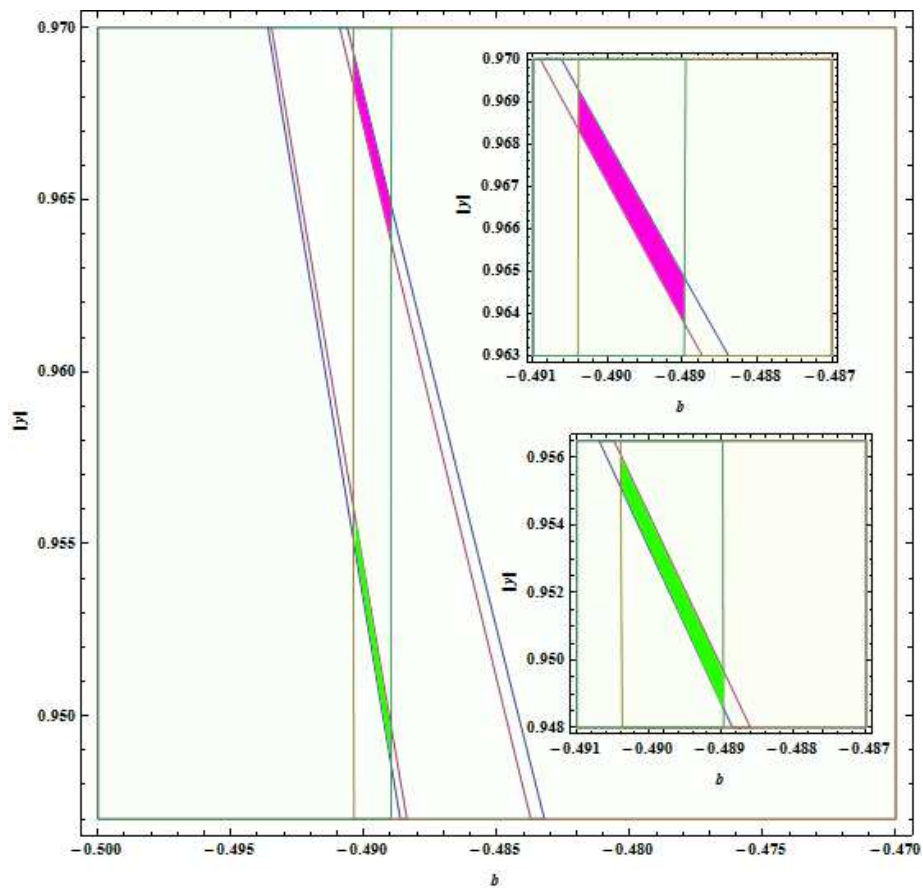}\caption{\label{fig.1} \small
   In this figure the whole region of the $|y|$-$b^{'}$ plane which is allowed by our model is shown.
   The overlap region of the experimental values for $\frac{\Delta
m_{21}^{2}}{\Delta m_{31}^{2}}$ and $\sin^{2}\theta_{13}$ with our
model are two tiny regions close to the top left corner of the
rectangular. In the zoomed box we have magnified the overlap
regions. As mentioned in the text, only the lower overlap region
is consistent with all of the experimental data.}
  \label{geometry}
\end{figure}
\end{center}

In this section we compare the experimental data with the results
obtained from our model. We do this by mapping two of the
constraints obtained from the experimental data onto our parameter
space, $b{'}$ and $y$, as shown in Figure (\ref{fig.1}). The
two most restricting experimental data come from the values of
$\frac{\Delta m_{21}^{2}}{\Delta m_{31}^{2}}$ and
$\sin^{2}\theta_{13}$. The overlap of $\frac{\Delta
m_{21}^{2}}{\Delta m_{31}^{2}}$ and $\sin^{2}\theta_{13}$ and our
model is restricted to two tiny regions close to the top left corner
of the parameter space.

First we check the consistency of the rest of the experimental data
with the overlap region. We find that the upper overlap region is barely ruled out by
the experimental value for $\sin^2\theta_{23}$, and the lower overlap region is
consistent with all the rest. Having determined the unique overlap region, we can not only determine
all of the parameters of our model, but also predict the masses of the neutrinos
and the phases.
The results for the parameters are as follows,

\begin{eqnarray}\label{eparameter1}
m_{0}&\approx&(3.3-3.7)10^{-2}eV ,\nonumber\\
a&\approx&-(1.71-1.93)10^{-2} eV,\nonumber\\
b&\approx&-(1.61-1.81)10^{-2}eV,\nonumber\\
B&\approx&(0.92-1.04)10^{-2}eV,\nonumber\\
\lambda&\approx&-(1.98-2.96)10^{-4}eV.
\end{eqnarray}
By comparing the values of these parameters we conclude that the
complexification part of the generalization of
$M_{FL}$, accomplished by adding imaginary components denoted by
B, could not be treated as a perturbation, while the second part
of the generalization, accomplished by adding $\lambda S^{123}$,
can indeed be treated as a perturbation. This is precisely what we
have done.
\begin{center}
\begin{table}[h2]
\centering
\begin{tabular}{|c|c|c|c|}
\hline Parameter & The exp. data    &The best exp. fit($\pm1\sigma$) &Predictions of our model
\\
\hline
$\Delta m_{21}^{2}(10^{-5}eV^{2})$  & $(7.12-8.20)$ &$(7.43-7.81)$&$(7.12-7.84)$  \\
\hline
$\Delta m_{31}^{2}(10^{-3}eV^{2})$ & $(2.26-2.77)$ &$(2.46-2.61)$&$ (2.23-2.80)$\\
\hline
$\sin^{2}\theta_{12}$   & $0.27-0.37$ & $0.30-0.336$ &$0.331-0.333$ \\
\hline
$\sin^{2}\theta_{13}$ & $0.016-0.037$ &$0.0281-0.0275$&$0.0218-0.0275$ \\
\hline
$\sin^{2}\theta_{23}$ & $0.39-0.64$ &$0.400-0.461$&$0.439-0.446$  \\
~&~&$0.573-0.635$&~\\
\hline
$\delta$ & $...$&$...$ & $59.21^{\circ}-59.34^{\circ}$ \\
\hline
$\sigma=-\rho$ & $...$&$...$&$59.44^{\circ}-59.51^{\circ}$ \\
\hline
$|J|$ & $...$&$...$&$0.029-0.033$ \\
\hline
$~$ & $...$&$...$&$m_1\approx(0.031625-0.035687) eV,$ \\
~&~&~&$\langle m_{\nu_e} \rangle\approx(0.033-0.037)~eV$\\
\cline{2-4}
masses & $...$&$...$&$m_2\approx(0.032732-0.036769)eV,$ \\
~&~&~&$\langle m_{\nu_\mu} \rangle\approx(0.043-0.048)~ eV$\\
\cline{2-4}
$~$ & $...$&$...$&$m_3\approx(0.056849-0.063819)eV,$ \\
~&~&~&$\langle m_{\nu_\tau} \rangle\approx(0.046-0.051)~ eV$\\
\hline
\end{tabular}
\caption{The available experimental data for neutrinos for the
case of normal mass hierarchy and the predictions of our model.
These predictions are obtained from our parameters as shown in
Eq.\,(\ref{eparameter1}). Specifically we have used $
b^{'}\approx-(0.489-0.4904)$ and
$y\approx(0.948932-0.956141).$}\label{h2} \label{tab:PPer}
\end{table}
\end{center}

Having determined all of the parameters of our model, we can
compare our results with the experimental data. In Table \ref{h2}
we state all of the relevant experimental data presented at
$3\sigma$ \cite{information1}, along with the predictions of our
model. As shown in Table \ref{h2} we have predictions for some
physical quantities for which no experimental data exist. We
should mention at this point that if we had chosen $S^{123}$ instead of $S^{132}$ to
break the $\mu-\tau$ symmetry and generate a nonzero value for
$\theta_{13}$, all of our final results and prediction would be
unaltered except for $J\rightarrow -J$.

An important experimental result for the sum of the three light neutrino masses
has just been reported by the \textit{Planck} measurements of the
cosmic microwave background (CMB)\cite{planck}, which is

\begin{equation}\label{eplanck}
\sum m_\nu<0.23 eV \text{(\textit{Planck}+WP+highL+BAO),}
\end{equation}
This sum in our model is $\sum m_\nu\approx(0.12-0.14)eV$,
which is consistent with the above constraint.
For the flavor eigenstates only the expectation values of the
masses can be calculated and they are obtained by the following relation,
\begin{equation}\label{enu}
\langle m_{\nu_{i} }\rangle=\sum_{j=1}^3|U_{ij}|^2|m_j|,
\end{equation}
where $i=e,\mu,~\text{and}~\tau$. Our predictions for these
quantities are also shown in Table \ref{h2}. The Majorana
neutrinos can violate lepton number, for example in neutrinoless
double beta decay $(0\nu \beta \beta)$ \cite{neutrinoless}. Such a
process has not yet been observed and an upper bound has been set
for the relevant quantity, {\em i.e.} $\langle m_{\nu_{ee}
}\rangle$. Results from the first phase of the KamLAND-Zen
experiment sets the following constraint $\langle m_{\nu_{ee}}
\rangle < (0.12 - 0.25)~eV $ at $90\%$ CL \cite{kamland}. Our
prediction for this quantity is $\langle m_{\nu_{ee}}
\rangle\approx(0.032-0.036)~eV$ which is consistent with the
result of kamLAND-Zen experiment. We also predict $\langle
m_{\nu_e} \rangle\approx(0.033-0.037)~ eV$, $\langle m_{\nu_\mu}
\rangle\approx(0.043-0.048)~ eV$ and $\langle m_{\nu_\tau}
\rangle\approx(0.046-0.051)~ eV$.

\section{Conclusions}

In this paper we propose a generalized Friedberg-Lee neutrino mass
model in which CP violation is possible. Our generalization
consists of two steps. First we add imaginary components to the
mass matrix, such that the resulting matrix is symmetric, magic
with zero trace. The results is that $\mu-\tau$ symmetry is
preserved and nonzero phases including $\delta$ are obtained. However, $\sin \theta_{13}$
remains zero. In the second stage we break the $\mu-\tau$ symmetry
by adding a real matrix which is the matrix representation of the
permutation group element $S^{(123)}$. This part is treated as a
perturbation and a nonzero $\theta_{13}$ is generated. The
combination of these two generalization then produces CP
violation. We use self consistency arguments and the complementary
ansatz to reduce our parameter space to two dimensions. Mapping
two of the experimental data, {\em i.e.} the allowed ranges of
$\sin^2\theta_{13}$ and $\frac{\Delta m^2_{21}}{\Delta m^2_{31}}$,
onto our allowed region of the parameter space almost pinpoints
the values of our parameters. We then check the consistency of all
other experimental data with the allowed range. We then predict
the mixing angles, the masses of neutrinos, the CP violation
parameters $\delta$, $\rho$, $\sigma$, and $J$. All of our
findings are shown in Table \ref{h2} along with the latest
experimental values. As is evident from Table \ref{h2}, all of our
predictions are within the allowed ranges determined by the
experiments and there is an excellent agreement between these
values. Our predictions for the neutrino masses and CP violation
parameters $\delta$ and $J$ are to tested by future experiments.

\section{Acknowledgement}
 We would like to thank the research office of the Shahid Beheshti University for financial
 support.


\end{document}